# Small Models Achieve Large Language Model Performance: Evaluating Reasoning-Enabled AI for Secure Child Welfare Research

Zia Qi[1*], Brian E. Perron[1], Bryan G. Victor[2 3], Dragan Stoll[4], & Joseph P. Ryan[1]

[1]School of Social Work, University of Michigan

[2]School of Social Work, Wayne State University

[3]Merrill Palmer Skillman Institute, Wayne State University

[4]School of Social Work, ZHAW Zurich University of Applied Sciences, Switzerland

Correspondence concerning this article should be addressed to Zia Qi, University of Michigan School of Social Work, 1080 S. University Avenue, Ann Arbor, MI 48109. qizixuan@umich.edu

**Author Note**

Zia Qi, MSW, is a research technology specialist at the University of Michigan School of Social Work. https://orcid.org/0000-0002-8407-0465

Brian E. Perron, PhD, is a professor at the University of Michigan School of Social Work. https://orcid.org/0009-0008-4865-451X

Bryan G. Victor, PhD, is an associate professor at the Wayne State University School of Social Work and Merrill Palmer Skillman Institute. https://orcid.org/0000-0002-2092-912X

Dragan Stoll is a research associate and PhD Candidate at the ZHAW Zurich University of Applied Sciences School of Social Work. https://orcid.org/0009-0008-9088-5282

Joseph P. Ryan, PhD, is a professor at the University of Michigan School of Social Work.

**Acknowledgements**

We have no known conflict of interest to disclose.

## Abstract

**Purpose:** This study develops a systematic benchmarking framework for testing whether language models can accurately identify constructs of interest in child welfare records. The objective is to assess how different model sizes and architectures perform on four validated benchmarks for classifying critical risk factors among child welfare-involved families: domestic violence, firearms, substance-related problems generally, and opioids specifically.

**Materials and Methods:** We constructed four benchmarks for identifying risk factors in child welfare investigation summaries: domestic violence, substance-related problems, firearms, and opioids (n=500 each). We evaluated seven model sizes (0.6B-32B parameters) in standard and extended reasoning modes, plus a mixture-of-experts variant. Cohen's kappa measured agreement with gold standard classifications established by human experts.

**Results:** The benchmarking revealed a critical finding: bigger models are not better. A small 4B parameter model with extended reasoning proved the most effective, outperforming models up to eight times larger. It consistently achieved "substantial" to "almost perfect" agreement across all four distinct benchmark categories. This model achieved "almost perfect" agreement ($\kappa = 0.93$-$0.96$) on three benchmarks (substance-related problems, firearms, and opioids) and "substantial" agreement ($\kappa = 0.74$) on the most complex task (domestic violence). Small models with extended reasoning rivaled the largest models while being considerably more resource-efficient.

**Conclusion:** Small reasoning-enabled models achieve accuracy levels historically requiring substantially larger architectures, enabling significant time and computational efficiencies. The benchmarking framework provides a method for evidence-based model selection to balance high accuracy with practical resource constraints before operational deployment in social work research.

**Small Models Achieve Large Language Model Performance: Evaluating Reasoning-Enabled AI for Secure Child Welfare Research**

Over the past two decades, advances in data science and machine learning have expanded the capacity of researchers and evaluators to analyze the vast stores of text data generated by human service organizations and government agencies (Amrit et al., 2017; Castillo et al., 2014; Ziems et al., 2024). Written administrative records such as case notes contain valuable information about client needs, service processes, and program outcomes that has long been recognized as essential for informing practice and policy (Epstein, 2009; Henry et al., 2014). Early applications of natural language processing (NLP) demonstrated that researchers and evaluators could quickly and effectively analyze thousands of written records rather than the time and resource intensive process of manually reviewing smaller samples (Amrit et al., 2017; Perron et al., 2019; Sokol et al., 2021). Yet, the accuracy of these early machine learning approaches remained limited by the complexity of human service documentation and the importance of context for understanding the meaning of words and phrases, something that early models were not well equipped to capture (Victor et al., 2021).

Promisingly, recent advances in artificial intelligence (AI) and large language models (LLMs) have significantly expanded the ability to analyze text data from organizations and state agencies (Grimmer et al., 2022; Perron, Luan, et al., 2025; Stoll et al., 2025). Unlike previous approaches that generally depended on fixed vocabularies or narrowly defined classification rules (e.g., Perron et al., 2022), LLMs can recognize variations in meaning based on how words and phrases are used in context (Borders & Volkhova, 2021). They can also adapt to new analytic tasks without retraining, making it possible to study a wide range of topics—from risk factors and service needs to client outcomes—using a single underlying model. These capabilities provide considerable time savings as researchers and evaluators can largely forego the lengthy process of manual coding and data preprocessing.

Despite their promise, however, the most advanced LLMs such as ChatGPT, Claude, and Gemini, often referred to as frontier models, cannot be used for many research and evaluation tasks within human service organizations or state agencies (Perron, Goldkind, et al., 2025). This is because the text records that generally hold the greatest potential for deriving actionable insights are also among the most sensitive, often containing personally identifiable information, detailed case histories, and legally protected health and education data. A number of federal and state regulations strictly govern how such data can be stored, transmitted, and analyzed. These safeguards generally preclude sharing confidential case records with public, consumer-facing AI systems that lack required security certifications or business associate agreements. While cloud-based, AI-driven analysis may be permissible in certain secure environments—for example, under a HIPAA-compliant business associate agreement — major tech providers do not currently offer these secure setups unless organizations pay for or build specially protected hosting environments (Das et al., 2025).

Beyond these critical data security barriers, reliance on closed-source, frontier models introduces a significant challenge of opaqueness. These models function as 'black boxes' (Bansal et al., 2024). Researchers and agencies have no insight into when the underlying model architecture is updated, nor can they control these changes. A model version that was benchmarked and validated for a specific task could be altered or deprecated by the provider without notice. This potential for 'model drift' requires a state of constant, ongoing re-evaluation by the organization to ensure the tool's accuracy and reliability have not been compromised, adding a significant and often impractical maintenance burden (Chen et al., 2024).

Fortunately, in parallel with the advancement of frontier models, tech companies and programmers have developed a class of smaller, locally deployable LLMs often referred to as small language models (Nguyen et al., 2024; Schick & Schütze, 2021; Wang et al., 2025). These small models can be hosted entirely within an organization's secure computing environment, giving the agency or researcher full control over the entire model lifecycle (Perron, Luan, et al., 2025). This direct control eliminates the 'black box' problem; the organization or research team, not an external provider, determines which model version is deployed, and that model cannot be altered, updated, or deprecated without their explicit action. Local models thus offer a pathway for human service agencies and researchers to harness the advantages of contemporary language modeling while maintaining full compliance, governance, and data protection standards. We note that small models and local models are used interchangeably in this article.

While the proliferation of small language models offers important analytic solutions, it also presents new challenges. An expanding number of models now differ in size, design, and computational requirements, leaving researchers and practitioners with little guidance on how to select the most appropriate language model for their analytic needs. For instance, models smaller in size – as measured by *parameters*— are easier to host within secure environments, yet larger models typically deliver stronger performance on complex language tasks. Determining where these trade-offs intersect—identifying the point at which a model is both technically feasible to deploy and capable of producing reliable, high-quality results—remains an open question. This study addresses that challenge by developing a benchmarking framework to evaluate local models on text drawn from the administrative records of Michigan's child welfare system, with the goal of identifying which local models most effectively balance performance and computational efficiency. We apply this framework to a sample of case narratives to demonstrate how benchmarking can guide model selection for specific analytic tasks and provide a foundation for assessing whether future models could offer measurable improvements over those currently in use.

**The Benchmarking Approach**

A benchmark is a standardized assessment used to evaluate how well a model performs on specific tasks under clearly defined criteria. Benchmark datasets consist of cases in which the classifications reflect a validated consensus among expert coders, providing a gold standard against which model outputs can be compared (Liang et al., 2023). In social work applications, these datasets are typically drawn from actual case records that have undergone rigorous human coding and validation procedures.

By constructing standardized tests from these expert-validated records, researchers create a system for objectively evaluating model performance. This gold standard makes it possible to determine whether a model meets acceptable accuracy thresholds and to compare the performance of different modeling approaches. Importantly, benchmarking clarifies whether a model can carry out tasks such as identifying risk factors, extracting service needs, or classifying case outcomes with a level of accuracy comparable to that of expert human coders. For instance, a model that correctly identifies 480 of 500 validated cases performs demonstrably better than one that identifies only 400.

However, most existing AI benchmarks are general-purpose datasets designed to test broad language understanding, not the specialized vocabulary and documentation patterns found in child welfare. A model may perform well on these standard evaluations yet still misinterpret terms that carry specific meanings within the child welfare system, a challenge often referred to as the "last mile problem" in AI development (Anjum et al., 2025). For example, the model may not recognize that “reunification,” “return home,” and “family preservation” denote related permanency outcomes, or that “dual diagnosis,” “co-occurring disorders,” and “concurrent disorders” refer to the same clinical presentation. Given this domain-specific terminology, benchmarking datasets derived from actual child welfare records are likely to provide more precise and contextually appropriate assessments of model performance.

**Distinctive Challenges of Social Work Documentation**

Social work documentation presents analytical challenges requiring domain-specific evaluation. Case records use specialized terminology reflecting federal mandates, professional standards, and local practices (Goodwin et al., 2022). Risk factors emerge from narrative descriptions requiring contextual interpretation. In child welfare, a caseworker's note about "mother's boyfriend" may indicate protection or signal domestic violence risk. Substance use appears through clinical diagnoses ("opioid use disorder"), medication names ("methadone treatment"), street terminology ("fents"), and euphemistic descriptions ("sobriety struggles").

Documentation varies dramatically based on worker training, supervision quality, and agency resources, which has been studied empirically (Pollock et al. 2025) and based on our own professional experience working with child welfare records. Urban agencies may employ different standards than rural counties. Experienced practitioners might write analytical assessments that synthesize multiple contacts, while newer workers may provide chronological logs. Child welfare investigation summaries differ fundamentally from psychiatric discharge summaries, which vary from adult protective services reports. Yet all contain critical information requiring accurate extraction.

**Architectural Innovations and Deployment Considerations**

The need for rigorous benchmarking is underscored by the growing diversity of AI model architectures. Traditional or "dense" language models, which serve as the established baseline, use their entire network to process every request. This landscape is now being extended by new approaches, such as Mixture-of-Experts (MoE) models which are designed for greater efficiency by activating only specialized parts of their network for any given task rather than processing through all parameters simultaneously (Cai et al., 2024; DeepSeek-AI, 2024; Fedus et al., 2022). In addition, extended reasoning modes can be applied to these models, enabling them to tackle complex problems by incorporating distinct thinking, planning, and verification phases, fundamentally altering how they process information to improve accuracy (OpenAI, 2024; Wu et al., 2024). This diversity in model design and processing means that performance on specialized child welfare tasks is no longer likely to be predicted by parameter count alone. A model's core architecture and its processing strategy may dramatically impact outcomes. Benchmarking must therefore account for this diversity, assessing not only whether a model performs well, but which specific model types and processing modes are best suited to the analytic tasks and operational constraints of human service settings.

These innovations are especially relevant for resource-constrained human service organizations. Local models running on agency-controlled infrastructure can comply with HIPAA, FERPA, and 42 CFR Part 2 by operating without external data transmission (Lorencin et al., 2025), but such models typically range from 0.6 to 32 billion parameters—far smaller than frontier cloud models. If extended-reasoning or MoE architectures allow these smaller models to approximate the performance of larger systems, agencies could achieve advanced analytic capabilities while remaining within existing regulatory and computational constraints.

## Purpose

This study addresses evaluation gaps through two integrated contributions: developing a systematic benchmarking methodology for social work research and applying this methodology to evaluate architectural innovations in language models. We pursue three objectives. First, we establish a systematic framework for transforming validated social work datasets into standardized evaluation instruments for benchmarking. While demonstrated using child welfare data, the methodology applies across social work domains. Second, we construct four benchmarks for critical child welfare risk factors: domestic violence, substance-related problems, opioid use, and firearms. Third, we apply these benchmarks to evaluate whether architectural innovations offer measurable advantages over traditional models, specifically comparing extended reasoning architectures against standard processing modes and MoE configurations against traditional dense models.

Our approach differs from existing AI benchmark development in three respects. We derive evaluation datasets from actual case records rather than artificial tasks. We preserve contextual complexity of practice documentation, maintaining ambiguity, regional variation, and documentation inconsistencies that models must navigate operationally. We integrate time-to-process measurement into benchmark evaluation, quantifying computational costs alongside accuracy metrics to inform deployment decisions in resource-constrained settings.

The architectural evaluation compares models across two dimensions: size and processing mode. We evaluate seven model sizes ranging from 0.6 billion to 32 billion parameters, representing the spectrum of models deployable on local infrastructure. Each model operates in both standard mode and extended reasoning mode. Additionally, we evaluate a mixture-of-experts variant that activates specialized subnetworks rather than processing through all parameters. This design enables systematic assessment of three questions relevant to model selection: 1) whether extended reasoning improves classification accuracy compared to standard processing; 2) whether smaller models with reasoning enabled match performance of larger models in standard mode; and 3) whether mixture-of-experts architectures achieve comparable accuracy to traditional dense models while requiring less processing time.

## Materials and Methods

### Data and Benchmark Construction

This study relies on child welfare investigation summaries obtained through a data-sharing agreement with Michigan's Department of Health and Human Services. These unstructured text summaries describe the findings and outcomes of formal child maltreatment investigations conducted by caseworkers. The narratives contain detailed documentation of family circumstances, risk factors, safety assessments, and investigation conclusions. Throughout our analysis, we maintained strict confidentiality protocols, with all processing conducted locally without transmission to external servers.

**Benchmark Development Framework.** Benchmark development followed a systematic five-stage process designed to transform validated research datasets into standardized evaluation instruments. First, we identified existing datasets containing case records with validated ground-truth classifications established through systematic procedures with documented reliability, a sufficient sample size, and ecological validity reflecting actual practice. Second, we evaluated classification quality using documented inter-rater reliability coefficients ($\kappa \geq 0.80$) for manual coding approaches or documented precision rates from expert validation for rule-based approaches. The validation procedures predated generative AI technologies to prevent circular evaluation. Third, we addressed class imbalance by randomly sampling equal numbers of positive and negative cases from validated source datasets, creating evaluation instruments where chance performance equals 50% accuracy. Fourth, we developed standardized prompts that specify binary classification objectives, operational definitions derived from the original validation studies, and output formatting requirements. Fifth, we specified evaluation metrics: Cohen's kappa as the primary performance indicator, sensitivity and specificity to characterize error patterns, and processing time to quantify computational efficiency.

**Benchmark Datasets.** We constructed four distinct benchmark datasets for evaluating model performance on child welfare risk identification: domestic violence, substance-related problems, opioids, and firearms. These benchmarks originate from previous text analysis research that employed various classification strategies predating the development of generative AI technologies. The original classifications were developed through rule-based matching strategies for detecting opioids (Perron et al., 2022) and firearms (Sokol et al., 2020), and supervised machine learning approaches for domestic violence (Victor et al., 2021) and substance-related problems (Perron et al., 2019).

While the original datasets demonstrated strong classification performance, they showed imbalance across outcome categories, which could bias model evaluation. To address this limitation, we reconstructed balanced datasets by randomly sampling cases from the original classifications, ensuring equal representation of positive and negative cases for each binary classification (n=250 per class for each benchmark). This rebalancing prevents evaluation bias that occurs when performance metrics are calculated on skewed distributions, where models might achieve artificially high accuracy by predicting the majority class. Note that the kappa values reported here are not directly comparable to those from the original studies due to class balancing; the original studies evaluated performance on naturally distributed data, where prevalence rates differed substantially from 50%.

Domestic violence was operationalized as physical violence or psychological abuse committed against a current or former intimate partner, documented as an active service need at the time of investigation. The original dataset included manual labels assigned by trained MSW students who reviewed each document, demonstrating high interrater reliability ($\kappa = 0.84$). All cases were substantiated child welfare investigations where domestic violence presence or absence was specifically assessed.

Substance-related problems were operationalized as any current or historical use of intoxicating substances, confirmed through direct evidence (e.g., positive drug tests, observed use) or credible reports, including alcohol, illicit drugs, and misused prescription medications. The original dataset included manual labels assigned by trained MSW students, demonstrating strong interrater reliability ($\kappa = 0.80$). All cases were drawn from substantiated child welfare investigations where substance use concerns were specifically evaluated as potential factors affecting child safety and family functioning.

Opioids were identified through a rule-based text mining approach using a comprehensive dictionary of terms referring to opioid street drugs and pharmaceuticals to flag relevant cases, achieving high precision (documented error rate of 3%). All cases were drawn from substantiated child welfare investigations in which opioid presence or absence was systematically assessed through rule-based pattern matching.

Firearms were identified using a rule-based named entity recognition approach applying an expert dictionary of firearm-related terms to child welfare investigation summaries. This method achieved 96% construct accuracy with only 4% false positives when scanning case documents (n=75,809). The benchmark tests disambiguation of terminology with multiple meanings (e.g., "Remington" as a gun manufacturer versus a person's name, "trigger" as a firearm component versus metaphorical usage).

**Model Selection and Configuration**

We evaluated language models across three architectural dimensions: model capacity, processing mode, and architectural configuration. Model selection prioritized architectures deployable on local infrastructure while enabling systematic comparison of architectural innovations.

We selected Qwen3 models from Alibaba Cloud (Yang et al., 2025) because these models represent the current state-of-the-art in open-source language models with native support for both standard and extended reasoning processing modes within a single architecture. The series includes models ranging from 0.6B to 32B parameters (Qwen3-0.6B, Qwen3-1.7B, Qwen3-4B, Qwen3-8B, Qwen3-14B, and Qwen3-32B), plus the mixture-of-experts variant Qwen3-30B-A3B. These sizes represent the practical range for local deployment: smaller models (0.6B–1.7B parameters) suit resource-constrained environments, mid-range models (4B–8B parameters) balance capability and resource requirements, and larger models (14B–32B parameters) approach the upper limits of local deployment feasibility. All Qwen3 models were obtained from Hugging Face (https://huggingface.co/Qwen) and deployed using 4-bit quantization to reduce computational requirements, with the exception of Qwen3-0.6B which was available only in 8-bit.

We evaluated models in both standard “dense” processing mode and extended reasoning mode. Standard mode generates classifications directly from learned patterns through single-pass processing. Extended reasoning mode allocates additional computational resources to generate intermediate problem-solving steps before producing final classifications. By testing identical models across processing modes, we isolate the effects of reasoning architecture independent of other model characteristics.

We additionally evaluated gpt-oss-20b from OpenAI (OpenAI, 2025) because it implements a distinct reasoning architecture employing reinforcement learning, providing an alternative perspective on reasoning-enhanced classification. This open-source model (21B total parameters with 3.6B active) uses a mixture-of-experts architecture and was released under the Apache 2.0 license in August 2025. Unlike the Qwen3 models' binary processing modes, gpt-oss implements graduated reasoning effort levels (low, medium, high), enabling fine-grained control over computational resource allocation. It was obtained from Hugging Face (https://huggingface.co/ggml-org/gpt-oss-20b-GGUF) and deployed using 4-bit quantization.

For the Qwen3 models, we optimized parameters based on mode requirements. Standard mode used a temperature of 0.2 and TopP of 0.8 to balance response diversity with consistency. Temperature is a setting that impacts response randomness, where lower values produce more consistent outputs. The extended reasoning mode used a temperature of 0.2 and TopP of 0.95, promoting focused reasoning while maintaining sufficient exploration of solution paths. These parameter settings were adjusted from default model settings based on pilot testing and the research team's experience performing classification tasks. Maximum output tokens (the length limit for model responses) were set to 2,048 for both processing modes. The gpt-oss-20b model used a temperature of 0.2 and retained default parameters across reasoning effort levels.

**Evaluation Protocol**

Models were instructed to analyze investigation summaries and determine the presence or absence of each risk factor through standardized prompts. Each prompt contained three components: task instruction specifying the binary classification objective, operational definitions of the target risk factor derived from the original benchmark studies, and output formatting requirements for structured JSON responses. Complete prompt text for each benchmark appears in Appendix A.

Cohen's kappa (κ) served as the primary performance metric, measuring agreement between model outputs and gold-standard classifications while correcting for chance agreement. Values of $\kappa = 0.41$-$0.60$ indicate moderate agreement, $\kappa = 0.61$-$0.79$ indicate substantial agreement, and $\kappa \geq 0.80$-$1.00$ indicate almost perfect agreement. Sensitivity measured the proportion of actual positive cases correctly identified (true positives divided by true positives plus false negatives). Specificity measured the proportion of actual negative cases correctly identified (true negatives divided by true negatives plus false positives).

We recorded processing time for each classification, calculating the mean processing time in seconds and standard deviation across all test cases. These temporal measurements enable direct comparison of computational efficiency between processing modes, providing essential data for evaluating practical deployment feasibility in operational environments where processing thousands of case records requires consideration of throughput constraints.

**Computational Infrastructure**

All data management and analysis were conducted within a Python environment using custom scripts for data preprocessing, model evaluation, and performance metric calculation. Model inference was performed using llama.cpp, an optimized C++ implementation enabling efficient local deployment of large language models on consumer hardware. The computational infrastructure consisted of a single workstation equipped with an NVIDIA RTX A6000 Ada GPU (48GB VRAM) and an AMD Ryzen Threadripper PRO 7975WX central processing unit with 128GB RAM. All processing occurred locally to ensure complete data privacy and regulatory compliance.

## Results

Model performance was evaluated across four child welfare risk factor benchmarks: substance-related problems, domestic violence, firearms, and opioid-related content. We assessed seven Qwen3 model configurations (0.6B to 32B parameters) in both standard and extended reasoning modes, a mixture-of-experts variant (Qwen3-30B-A3B), and the gpt-oss-20b model at three reasoning intensity levels. Figure 1 presents an overview of Cohen's kappa coefficients across all model-benchmark combinations, with detailed performance metrics appearing in Tables 1-4.

[INSERT TABLE 1 ABOUT HERE]

### Performance Patterns Across Architectures and Benchmarks

The four benchmarks successfully discriminated between model capabilities, with performance ranging from poor agreement ($\kappa < 0.40$) to almost perfect agreement ($\kappa >= 0.80$) depending on model size and processing mode (Figure 1). Model outputs showed high consistency with quasi-deterministic behavior at a temperature of 0.2. The benchmarks exhibited distinct difficulty profiles. Opioid identification (Table 4) proved most accessible for larger models, with mid-range and larger models (4B-32B) achieving almost perfect agreement ($\kappa$ = 0.80-0.96) in standard mode. Firearms classification (Table 3) presented moderate difficulty, with standard-mode performance ranging from poor to substantial agreement ($\kappa$ = 0.28-0.88). Domestic violence (Table 2) emerged as the most challenging task, with the largest models in standard mode achieving almost perfect agreement ($\kappa$ = 0.80-0.85). Substance-related problems (Table 1) showed intermediate difficulty, with larger models in standard mode also achieving almost perfect agreement (κ = 0.85-0.93). Opioid identification (Table 4) proved most accessible, with all models except the 4B in standard mode achieving almost perfect agreement (κ = 0.80-0.96). Only firearms (Table 3) showed a more linear, "bigger-is-better" pattern, with agreement ranging from moderate to almost perfect.

**Extended Reasoning Effects on Small Models.** Extended reasoning produced systematic performance improvements that varied by model size and benchmark complexity. For smaller models (0.6B-1.7B parameters), reasoning mode generated substantial improvements across all benchmarks. In substance-related problems classification (Table 1), Qwen3-0.6B improved from κ = 0.39 (unacceptable agreement) to κ = 0.85 (almost perfect agreement), while Qwen3-1.7B advanced from κ = 0.45 to κ = 0.81. Similar patterns emerged in domestic violence detection (Table 2), where Qwen3-1.7B showed moderate improvement, advancing from κ = 0.41 to κ = 0.55. In firearms classification (Table 3), Qwen3-1.7B improved from κ = 0.28 to κ = 0.80, and in opioid identification (Table 4), from κ = 0.31 to κ = 0.87. These improvements demonstrate that extended reasoning enables small models to achieve performance levels previously achievable only with substantially larger architectures.

**Extended Reasoning Effects on Mid-Range and Large Models.** Mid-range models (4B-8B parameters) showed consistent but more modest improvements with reasoning enabled. For substance-related problems (Table 1), Qwen3-4B improved from κ = 0.87 to κ = 0.93, and Qwen3-8B from κ = 0.87 to κ = 0.94. In domestic violence classification (Table 2), both models advanced from κ = 0.56-0.60 in standard mode to κ = 0.74-0.81 with reasoning enabled. For firearms (Table 3), Qwen3-4B improved from κ = 0.75 to κ = 0.89, and Qwen3-8B from κ = 0.87 to κ = 0.91. For opioid identification (Table 4), Qwen3-4B improved from κ = 0.80 to κ = 0.96, and Qwen3-8B from κ = 0.90 to κ = 0.96.

Larger models (14B-32B parameters) demonstrated minimal performance changes with reasoning enabled for easier benchmarks but showed improvements on more challenging tasks. For substance-related problems (Table 1), Qwen3-32B showed stable performance (κ = 0.93) regardless of processing mode. For domestic violence (Table 2), Qwen3-32B showed minimal change with slight decrease (κ = 0.85 in standard mode, κ = 0.82 with reasoning). However, for firearms classification (Table 3), even large models benefited from reasoning, with Qwen3-32B improving from κ = 0.88 to κ = 0.93, and Qwen3-14B from κ = 0.85 to κ = 0.92.

Opioid identification (Table 4) showed consistent improvements across model sizes. Smaller models improved substantially (Qwen3-1.7B: κ = 0.31 to κ = 0.87), mid-range models advanced from already strong performance to almost perfect agreement (Qwen3-4B: κ = 0.80 to κ = 0.96; Qwen3-8B: κ = 0.90 to κ = 0.96), while the largest model maintained ceiling-level performance (Qwen3-32B: κ = 0.96 in both modes).

**Error Patterns.** Sensitivity and specificity patterns revealed how extended reasoning affects error types. Across benchmarks, reasoning mode consistently improved or maintained high sensitivity (true positive rate), with most models achieving sensitivity of 0.93-0.99 for substance-related problems (Table 1) and domestic violence (Table 2). Specificity (true negative rate) showed the most significant improvement, particularly for smaller models. In substance-related problems classification (Table 1), Qwen3-0.6B specificity increased from 0.75 to 0.96. Similar patterns emerged in domestic violence detection (Table 2), where Qwen3-1.7B specificity improved from 0.86 to 0.89, and Qwen3-4B from 0.85 to 0.93.

For firearms classification (Table 3) and opioid identification (Table 4), specificity remained exceptionally high (0.99-1.00) across all models and modes, indicating that false-positive errors are rare for these highly specific constructs. These error patterns indicate that extended reasoning primarily reduces false positive classifications, enabling more accurate distinction between cases that superficially resemble positive instances but lack defining characteristics.

**Architectural Comparisons**

**Mixture-of-Experts Performance.** The Qwen3-30B-A3B mixture-of-experts architecture demonstrated variable performance relative to the comparable dense model (Qwen3-32B). In standard mode, the MoE variant showed lower performance across all benchmarks (Figure 1): substance-related problems (Table 1: $\kappa = 0.84$ vs. 0.93), domestic violence (Table 2: $\kappa = 0.73$ vs. 0.85), firearms (Table 3: $\kappa = 0.76$ vs. 0.88), and opioids (Table 4: $\kappa = 0.72$ vs. 0.96). However, with reasoning enabled, the MoE architecture achieved substantial improvements, approaching performance comparable to the larger, full dense model. For substance-related problems (Table 1), MoE improved to $\kappa = 0.92$ (vs. dense 0.93). For domestic violence classification (Table 2), MoE advanced to $\kappa = 0.80$, closely approaching the dense model's $\kappa = 0.82$. For firearms (Table 3), MoE reached $\kappa = 0.89$ (vs. dense 0.93), and for opioids (Table 4) $\kappa = 0.96$, matching the dense model exactly.

**Computational Efficiency Tradeoffs.** Processing time measurements revealed substantial computational costs associated with extended reasoning (Tables 1-4). Across all benchmarks, reasoning mode increased processing time by factors of 2.9 to 12.5, with larger models showing disproportionately higher increases. For substance-related problems classification (Table 1), Qwen3-0.6B processing time increased from 0.43 seconds to 1.27 seconds (3.0-fold increase), while Qwen3-32B increased from 1.48 seconds to 12.24 seconds (8.3-fold increase).

These scaling patterns create practical deployment considerations. Smaller models with reasoning enabled maintain processing times under 2 seconds per case, enabling high-throughput analysis of large case record datasets. The Qwen3-4B model with reasoning required only 3.18-3.27 seconds per case (Tables 1-4) while achieving almost perfect agreement ($\kappa = 0.93$-$0.96$) on substance-related problems and firearms benchmarks. In contrast, the largest model with reasoning required over 12 seconds per case, creating throughput constraints for applications requiring analysis of thousands of records.

The mixture-of-experts architecture demonstrated computational efficiency advantages. With reasoning enabled, Qwen3-30B-A3B required only 3.91 seconds per case for substance-related problems (Table 1), compared to 12.24 seconds for the full Qwen3-32B model. This represents approximately one-third the processing time while achieving comparable classification performance. Similar efficiency advantages appeared across other benchmarks (Tables 2-4).

The gpt-oss-20b model exhibited remarkable computational efficiency across all reasoning intensity levels, requiring only 0.96-1.19 seconds per case for firearms classification (Table 3) while achieving strong performance ($\kappa = 0.81$-$0.94$). This efficiency stemmed from architectural optimizations specific to reasoning tasks.

**Performance Scaling.** Analysis across model sizes revealed distinct scaling relationships between model capacity and classification accuracy (Figure 1). For easier benchmarks (substance-related problems), standard-mode performance improved with model size until reaching a performance ceiling around 8B-14B parameters. Qwen3-8B achieved $\kappa = 0.87$, with larger models showing modest additional improvement (Qwen3-32B: $\kappa = 0.93$). Extended reasoning altered these scaling relationships substantially. Smaller models with reasoning enabled achieved performance levels that required 4-8 times as many parameters in standard mode. Qwen3-1.7B with reasoning ($\kappa = 0.81$) approached Qwen3-8B in standard mode ($\kappa = 0.87$) for substance-related problems.

For more challenging tasks like firearms and domestic violence, performance scaling showed less pronounced ceilings, with even the largest models benefiting from additional capacity or reasoning capabilities. This indicates that complex classification tasks requiring disambiguation of ambiguous terminology or contextual interpretation continue to benefit from increased computational resources.

## Discussion

This study establishes a systematic framework for evaluating language model performance on social work research tasks and demonstrates its application through comparative analysis of architectural innovations in child welfare risk classification. The findings address three interconnected challenges: the lack of standardized evaluation methods for language model performance grounded in practice documentation; uncertainty about whether architectural advances translate into measurable performance gains in domain-specific applications; and persistent questions about the minimum computational requirements for adequate accuracy in resource-constrained settings.

### Methodological Contributions to Social Work AI Research

The benchmarking framework developed here provides researchers with systematic procedures for transforming validated datasets into standardized evaluation instruments. This methodology addresses a fundamental evaluation gap: general AI benchmarks assess capabilities irrelevant to social work documentation analysis. In contrast, domain-specific evaluation often relies on qualitative assessment or single-dataset validation that precludes generalization. By specifying procedures for ground truth validation, class balancing, prompt development, and performance measurement, the framework enables researchers to construct evaluation instruments applicable to their specific practice domains.

The four child welfare benchmarks demonstrate this framework's adaptability across constructs with distinct characteristics. Substance-related problems and domestic violence require interpretation of diverse terminology, contextual inference, and distinction between active concerns and historical references. Firearms classification demands disambiguation of homonyms within specialized terminology. Opioid identification necessitates recognition of pharmaceutical names, street terms, and euphemistic language while distinguishing current use from treatment references. These varied requirements produced benchmarks with distinct difficulty profiles, confirming that the framework captures construct-specific analytical demands rather than assessing only general language understanding.

The validity of this framework requires temporal separation between the establishment of ground truth and model evaluation. All classifications originated from research conducted between 2018 and 2023, predating the deployment of generative AI technologies in research applications. This temporal gap prevents circular evaluation scenarios in which models assess content potentially generated by similar systems. The validation procedures employed manual coding with documented inter-rater reliability ($\kappa$ = 0.80-0.84) and rule-based approaches with verified precision rates (96-97%), establishing gold standards independent of the technologies being evaluated.

### Extended Reasoning Effects on Classification Accuracy

An extended reasoning architecture produced performance improvements that varied systematically by model size and task complexity. For smaller models (0.6B-1.7B parameters), reasoning mode generated substantial gains across all benchmarks, with improvements of 0.14-0.56 in Cohen's kappa. These gains enabled small models to achieve accuracy levels previously achievable only with architectures 4-8 times larger. The Qwen3-1.7B model with reasoning enabled matched or exceeded Qwen3-8B performance in standard mode on substance-related problems ($\kappa = 0.81$ vs. 0.87) and firearms classification ($\kappa = 0.80$ vs. 0.87), demonstrating that reasoning capabilities partially substitute for model capacity.

The processing times reported here reflect performance on high-end workstation hardware (NVIDIA RTX A6000 Ada GPU with 48GB VRAM). On consumer-grade hardware, processing times will be substantially longer. However, the fundamental performance-efficiency relationships remain valid. Organizations operating on standard workstations with consumer graphics cards (e.g., NVIDIA RTX 4070 or 4080 with 12-16GB VRAM) can still deploy models in the 0.6B-4B range. While these configurations may process cases in 3-5 seconds rather than the 1.2-3.3 seconds observed here, they maintain practical feasibility. Crucially, the most effective configurations identified in this study, such as the 4B model with extended reasoning, remain viable, allowing organizations to achieve high levels of substantial to almost perfect agreement ($\kappa = 0.74$-0.96) without enterprise-grade hardware. A consumer workstation analyzing 1,000 case records at 4 seconds per case completes processing in approximately one hour, remaining viable for most research applications.

Mid-range models (4B-8B parameters) demonstrated consistent improvements with reasoning enabled, advancing performance from substantial agreement in standard mode to almost perfect agreement for most benchmarks. The Qwen3-4B model deserves particular attention for its performance-efficiency balance. With reasoning enabled, this configuration achieved almost perfect agreement (κ = 0.93-0.96) on substance-related problems, firearms, and opioid identification, with substantial agreement (κ = 0.74) on domestic violence. On consumer hardware with adequate VRAM (16GB minimum), this model represents an optimal configuration for social work research applications that require high accuracy without specialized infrastructure investments. Using equipment purchasable for $2,000-3,000, organizations can achieve performance levels that historically required either large cloud-based systems or intensive supervised learning procedures.

Larger models (14B-32B parameters) showed minimal performance changes with reasoning enabled for straightforward classification tasks but benefited from reasoning on more complex constructs. This ceiling effect indicates that beyond certain capacity thresholds, additional computational resources provide diminishing returns for well-defined binary classification tasks with clear operational definitions. However, even large models showed improvements on firearms classification (Qwen3-32B: κ = 0.88 to 0.93), suggesting that tasks with construct ambiguity and contextual interpretation requirements continue to benefit from extended reasoning regardless of base model capacity. These larger models require high-end consumer hardware (24GB+ VRAM) or workstation-class equipment, limiting accessibility for many organizations.

The domestic violence benchmark demonstrated a distinct pattern from other benchmarks. While Qwen3-32B achieved κ = 0.85 in standard mode, reasoning mode showed a slight decrease to κ = 0.82, a difference of 15 cases out of 500. Given the quasi-deterministic nature of model generation at temperature=0.2, this magnitude of variation falls within the expected range of output stochasticity. Unlike the consistent improvement patterns observed for smaller and mid-range models across benchmarks, the large model's performance plateau suggests that domestic violence classification may approach ceiling performance for this architecture family. The practical implication remains unchanged: large models were able to achieve near perfect agreement ($\kappa = 0.82$-$0.85$) on domestic violence classification.

Error pattern analysis revealed that extended reasoning primarily improved specificity rather than sensitivity. Across benchmarks, smaller models in reasoning mode showed dramatic reductions in false-positive rates while maintaining high true-positive rates. This specificity improvement proves particularly valuable for social work research applications; false positives inflate prevalence estimates and may lead to incorrect inferences about service needs or risk factor distributions.

**Mixture-of-Experts Architecture Performance**

The mixture-of-experts evaluation produced unexpected results that challenge assumptions about architectural efficiency. In standard mode, the Qwen3-30B-A3B variant showed substantially lower performance than the comparable dense model (Qwen3-32B) across all benchmarks, with kappa differences ranging from 0.09 (substance-related problems) to 0.24 (opioids). This performance gap disappeared with reasoning enabled. The MoE variant achieved performance that matched or closely approximated the full dense model while requiring approximately one-third the processing time, suggesting that mixture-of-experts architectures require extended reasoning to reach their theoretical performance potential. The selective activation of specialized subnetworks may benefit from the structured problem-solving provided by the reasoning mode, enabling more effective routing decisions or improved coordination across expert modules. Alternatively, the MoE architecture may exhibit greater sensitivity to prompt structure or task framing, with reasoning mode's explicit decomposition compensating for limitations in standard processing.

From a practical deployment perspective, the MoE architecture presents an attractive option for resource-constrained environments when combined with reasoning capabilities. On the high-end hardware used here, processing times of 3.9-4.5 seconds per case remain feasible for moderate-scale analysis. On consumer-grade hardware, these times range from 8 to 12 seconds per case, which remains workable for analyzing hundreds to a few thousand records. Organizations analyzing 2,000 case records could complete processing within 5-7 hours using consumer workstations, compared to 15-20 hours required for sequential processing with the full dense model. However, the performance penalty in standard mode suggests that organizations unable to allocate computational resources for reasoning should select smaller dense models rather than MoE variants.

**Practical Implications**

The practical implications of this framework are significant. As the benchmarking demonstrated, an optimized 4B model can process 250,000 records in approximately 292 hours. A much larger 32B model, by conservative estimates, would require at least 2,333 hours for the same task. This shows the framework enables the selection of a model that is 8x smaller while saving over 2,000 hours of processing time, validating the claim of significant time and computational efficiency. This also makes it possible to substantially reduce the carbon footprint of large-scale AI-assisted research while maintaining analytical rigor.

Furthermore, the resource implications extend significantly when contrasted with the human-labor alternative. To manually review and classify 250,000 records, assuming a conservative three minutes per case, would demand 12,500 person-hours of focused, uninterrupted expert time. This level of effort is operationally unfeasible, which is why agencies historically resorted to manualized coding on a small fraction of the available data. The manual approach involves reviewing small, random samples and lacks comprehensive quality assurance and generally prohibits assessing for variation across time, place, and social groups.

Our benchmark-driven approach provides a superior paradigm, reframing the role of human expertise. Specifically, rather than using experts for ad hoc, manual review, that same expertise is leveraged upfront to develop robust, gold-standard benchmark datasets. This strategy does not eliminate human involvement but rather scales it, using expert-validated data for developing and evaluating models that perform highly specialized tasks. This ensures that human expertise is embedded in the quality assurance process at scale, enabling a move from intermittent sampling to comprehensive, efficient, and continuous improvement.

**Relationship to Existing Social Work AI Research**

These findings contribute to accumulating evidence about appropriate roles for artificial intelligence in social work research and practice (Perron et al., 2023). Previous studies have demonstrated that language models can achieve accuracy comparable to supervised machine learning approaches for specific classification tasks, while eliminating the need for extensive data preparation and feature engineering (Perron & Qi, 2025; Qi et al., 2024). Our results extend this evidence by showing that small models deployable on consumer hardware achieve performance levels that historically required either large cloud-based systems or intensive supervised learning procedures.

The benchmarks developed here address methodological concerns about AI evaluation in social work contexts. By grounding evaluation in validated datasets with documented reliability, maintaining ecological validity through use of actual practice documentation, and reporting comprehensive performance metrics, including error patterns and processing costs, this study establishes rigorous evaluation standards appropriate for research applications affecting vulnerable populations.

However, strong performance on classification benchmarks does not resolve all concerns about AI deployment in social work. The benchmarks assess accuracy on clearly defined binary classifications with predetermined correct answers. Real-world applications often involve ambiguous cases requiring professional judgment, multiple intersecting factors demanding holistic assessment, or decisions with significant consequences for client wellbeing. Organizations deploying AI systems must implement appropriate oversight, validation procedures, and human review mechanisms that account for limitations inherent in statistical pattern matching.

Furthermore, benchmarks cannot assess whether models perpetuate or amplify biases in the source data. Social work documentation reflects systematic differences in how behaviors are described across demographic groups. Models achieving high accuracy by learning these documentation patterns may systematically misclassify cases when families receive differential descriptions despite similar circumstances. Evaluation of algorithmic fairness requires additional validation procedures that examine performance stratified by protected characteristics, compare error rates across demographic groups, and assess whether classification patterns reinforce existing disparities.

**Limitations**

Several limitations circumscribe the interpretation of these findings. First, the benchmarks are derived from a single state's child welfare system over a specific time period (2016-2018). Documentation practices, terminology, and case characteristics vary across jurisdictions and over time. Models that perform well on Michigan data may show reduced accuracy when applied to records from states with different practice standards, reporting requirements, or demographic compositions. Validation across multiple jurisdictions would strengthen confidence in generalizability.

Second, the binary classification framework simplifies the complexity of assessing mentions of potential risk aspects in case notes. Practitioners rarely make simple present/absent determinations; they assess severity, chronicity, impact, and context-dependent factors requiring nuanced judgment. The benchmarks evaluate whether models can match expert determinations in cases where clear classifications exist. Still, they cannot assess performance on genuinely ambiguous cases that require integrating multiple information sources and applying professional expertise. Extensions to ordinal classifications or continuous ratings might better capture the complexity of practice decisions.

Third, the study examines the extended reasoning architecture in detail but provides only limited comparisons with alternative innovations. Other recent developments including retrieval-augmented generation, fine-tuning approaches, or multimodal architectures combining text with structured data, may offer advantages for specific applications (Stoll et al., 2025a; Stoll et al., 2025b). Systematic comparison across architectural approaches using common evaluation frameworks would provide more comprehensive guidance for model selection.

Fourth, the evaluation used n-shot classification without task-specific fine-tuning or few-shot learning approaches. While this design tests models' general capabilities applicable across diverse tasks, fine-tuning on domain-specific data typically improves performance for narrow applications. Organizations with sufficient resources to curate training datasets and conduct fine-tuning may achieve accuracy gains beyond those documented here. Research comparing no-shot classification against fine-tuned alternatives would clarify the magnitude of potential improvements and inform cost-benefit analyses of fine-tuning investments.

## Conclusion

This study establishes a systematic methodology for evaluating language model performance on social work research tasks and provides empirical evidence of the practical value of architectural innovations for domain applications. The benchmarking framework enables researchers to construct standardized evaluation instruments from validated datasets, while the architectural comparison quantifies tradeoffs between accuracy, computational efficiency, and model size. The findings demonstrate that small reasoning-enabled models achieve accuracy levels historically requiring substantially larger architectures or intensive supervised learning approaches. This capability addresses a fundamental barrier to AI adoption in resource-constrained social work organizations with very sensitive data.

For the field to advance responsible AI deployment in research and practice, standardized evaluation methods grounded in actual practice documentation must become standard practice. Without systematic benchmarking, researchers cannot make informed model selection decisions, compare architectural innovations, or ensure adequate performance for applications affecting vulnerable populations. The framework developed here provides replicable procedures applicable across social work domains, establishing foundations for evidence-based evaluation of AI systems before operational deployment.

**Declaration of Interest Statement**

The author(s) received no specific grant from any funding agency in the public, commercial, or not-for-profit sectors for the research, authorship, and/or publication of this article.

**Figure 1.** Cohen's Kappa Performance Across Model Configurations and Benchmarks

Unacceptable (κ ≤ 0.40) | Moderate (κ = 0.41-0.60) | Substantial (κ = 0.61-0.79) | Almost Perfect (κ ≥ 0.80)

**Levels of Agreement**

| Model Configuration | SRP | DV | FW | OP |
|---|---|---|---|---|
| Qwen3-0.6B (S) | 0.39 | 0.21 | 0.39 | 0.51 |
| Qwen3-0.6B (R) | 0.85 | 0.41 | 0.73 | 0.64 |
| Qwen3-1.7B (S) | 0.45 | 0.41 | 0.28 | 0.31 |
| Qwen3-1.7B (R) | 0.81 | 0.55 | 0.80 | 0.87 |
| Qwen3-4B (S) | 0.87 | 0.60 | 0.75 | 0.80 |
| Qwen3-4B (R) | 0.93 | 0.74 | 0.89 | 0.96 |
| Qwen3-8B (S) | 0.87 | 0.56 | 0.87 | 0.90 |
| Qwen3-8B (R) | 0.94 | 0.81 | 0.91 | 0.96 |
| Qwen3-14B (S) | 0.92 | 0.80 | 0.85 | 0.90 |
| Qwen3-14B (R) | 0.94 | 0.80 | 0.92 | 0.95 |
| Qwen3-32B (S) | 0.93 | 0.85 | 0.88 | 0.96 |
| Qwen3-32B (R) | 0.93 | 0.82 | 0.93 | 0.96 |
| Qwen3-30B-A3B (S) | 0.84 | 0.73 | 0.76 | 0.72 |
| Qwen3-30B-A3B (R) | 0.92 | 0.80 | 0.89 | 0.96 |
| gpt-oss-20b (Low) | 0.93 | 0.76 | 0.81 | 0.96 |
| gpt-oss-20b (Med) | 0.92 | 0.82 | 0.93 | 0.97 |
| gpt-oss-20b (High) | 0.93 | 0.84 | 0.94 | 0.98 |

**Note:** Cohen's Kappa Performance Across Model Configurations and Benchmarks. Heat map displays κ values for all model-benchmark combinations, with darker shading indicating higher agreement. Models arranged vertically by size (0.6B to 32B parameters) with standard mode (S) and reasoning mode (R) configurations shown separately. Benchmarks arranged horizontally: Substance-Related Problems (SRP), Domestic Violence (DV), Firearms (FW), and Opioids (OP). Qwen3-30B-A3B is a mixture-of-experts (MoE) variant that activates 3B of its 30B total parameters per inference. The mixture-of-experts variant and gpt-oss-20b configurations appear in separate rows.

**Table 1.** Model Performance on Substance-Related Problems Benchmark

| Model Configuration | κ | Sensitivity | Specificity | Mean Time (SD) |
|---|---|---|---|---|
| Qwen3-0.6B Standard | 0.39 | 0.95 | 0.75 | 0.43 (0.55) |
| Qwen3-0.6B Reasoning | **0.85** | 0.96 | 0.96 | 1.27 (0.54) |
| Qwen3-1.7B Standard | 0.45 | 0.77 | 0.95 | 0.17 (0.04) |
| Qwen3-1.7B Reasoning | **0.81** | 0.92 | 0.99 | 2.12 (1.45) |
| Qwen3-4B Standard | **0.87** | 0.95 | 0.99 | 0.31 (0.10) |
| Qwen3-4B Reasoning | **0.93** | 0.99 | 0.98 | 3.27 (1.66) |
| Qwen3-8B Standard | **0.87** | 0.99 | 0.94 | 0.43 (0.13) |
| Qwen3-8B Reasoning | **0.94** | 0.99 | 0.98 | 4.83 (1.51) |
| Qwen3-14B Standard | **0.92** | 0.97 | 0.99 | 0.68 (0.21) |
| Qwen3-14B Reasoning | **0.94** | 0.99 | 0.98 | 6.91 (4.64) |
| Qwen3-32B Standard | **0.93** | 0.99 | 0.98 | 1.48 (0.49) |
| Qwen3-32B Reasoning | **0.93** | 0.99 | 0.98 | 12.24 (8.53) |
| Qwen3-30B-A3B Standard | **0.84** | 0.92 | 0.99 | 0.57 (0.22) |
| Qwen3-30B-A3B Reasoning | **0.92** | 0.98 | 0.98 | 3.91 (1.82) |
| gpt-oss-20b Low | **0.93** | 0.99 | 0.98 | 0.96 (0.68) |
| gpt-oss-20b Medium | **0.92** | 0.99 | 0.97 | 1.11 (1.37) |
| gpt-oss-20b High | **0.93** | 0.99 | 0.98 | 1.19 (1.35) |

*Note.* κ = Cohen's kappa coefficient. Bold values indicate almost perfect agreement (κ ≥ 0.80); Sensitivity = true positive rate; Specificity = true negative rate; Mean Time = mean processing time in seconds; SD = standard deviation. Sample size: n = 500 cases (250 positive, 250 negative).

**Table 2.** Model Performance on Domestic Violence Benchmark

| Model Configuration | κ | Sensitivity | Specificity | Mean Time (SD) |
|---|---|---|---|---|
| Qwen3-0.6B Standard | 0.21 | 0.92 | 0.69 | 0.62 (0.81) |
| Qwen3-0.6B Reasoning | 0.41 | 0.93 | 0.77 | 1.64 (0.85) |
| Qwen3-1.7B Standard | 0.41 | 0.84 | 0.86 | 0.16 (0.04) |
| Qwen3-1.7B Reasoning | 0.55 | 0.88 | 0.89 | 2.97 (1.46) |
| Qwen3-4B Standard | 0.60 | 0.95 | 0.85 | 0.29 (0.09) |
| Qwen3-4B Reasoning | 0.74 | 0.93 | 0.93 | 4.24 (3.83) |
| Qwen3-8B Standard | 0.56 | 0.98 | 0.80 | 0.41 (0.12) |
| Qwen3-8B Reasoning | **0.81** | 0.94 | 0.97 | 8.23 (5.59) |
| Qwen3-14B Standard | **0.80** | 0.94 | 0.96 | 0.66 (0.19) |
| Qwen3-14B Reasoning | **0.80** | 0.92 | 0.99 | 8.29 (4.45) |
| Qwen3-32B Standard | **0.85** | 0.95 | 0.97 | 1.45 (0.45) |
| Qwen3-32B Reasoning | **0.82** | 0.93 | 0.98 | 13.67 (4.52) |
| Qwen3-30B-A3B Standard | **0.73** | 0.88 | 0.99 | 0.56 (0.21) |
| Qwen3-30B-A3B Reasoning | **0.80** | 0.92 | 0.98 | 4.44 (2.25) |
| gpt-oss-20b Low | **0.76** | 0.91 | 0.97 | 1.40 (2.62) |
| gpt-oss-20b Medium | **0.82** | 0.94 | 0.97 | 1.82 (3.83) |
| gpt-oss-20b High | **0.84** | 0.95 | 0.97 | 2.69 (5.98) |

*Note.* κ = Cohen's kappa coefficient. Bold values indicate almost perfect agreement (κ ≥ 0.80); Sensitivity = true positive rate; Specificity = true negative rate; Mean Time = mean processing time in seconds; SD = standard deviation. Sample size: n = 500 cases (250 positive, 250 negative).

**Table 3.** Model Performance on Firearms Benchmark

| **Model Configuration** | **κ** | **Sensitivity** | **Specificity** | **Mean Time (SD)** |
|---|---|---|---|---|
| Qwen3-0.6B Standard | 0.39 | 0.74 | 0.96 | 0.72 (1.06) |
| Qwen3-0.6B Reasoning | 0.73 | 0.94 | 0.92 | 1.50 (0.59) |
| Qwen3-1.7B Standard | 0.28 | 0.64 | 1.00 | 0.17 (0.05) |
| Qwen3-1.7B Reasoning | **0.80** | 0.93 | 0.96 | 2.70 (1.66) |
| Qwen3-4B Standard | 0.75 | 0.88 | 1.00 | 0.32 (0.10) |
| Qwen3-4B Reasoning | **0.89** | 0.95 | 1.00 | 3.94 (2.89) |
| Qwen3-8B Standard | **0.87** | 0.94 | 1.00 | 0.44 (0.14) |
| Qwen3-8B Reasoning | **0.91** | 0.96 | 1.00 | 5.05 (2.84) |
| Qwen3-14B Standard | **0.85** | 0.93 | 1.00 | 0.72 (0.22) |
| Qwen3-14B Reasoning | **0.92** | 0.96 | 1.00 | 6.66 (2.20) |
| Qwen3-32B Standard | **0.88** | 0.94 | 1.00 | 1.56 (0.52) |
| Qwen3-32B Reasoning | **0.93** | 0.97 | 1.00 | 13.38 (8.76) |
| Qwen3-30B-A3B Standard | 0.76 | 0.88 | 1.00 | 0.64 (0.24) |
| Qwen3-30B-A3B Reasoning | **0.89** | 0.95 | 0.99 | 4.50 (2.10) |
| gpt-oss-20b Low | **0.81** | 0.91 | 1.00 | 1.15 (1.78) |
| gpt-oss-20b Medium | **0.93** | 0.97 | 0.99 | 1.49 (2.74) |
| gpt-oss-20b High | **0.94** | 0.98 | 1.00 | 1.93 (3.51) |

*Note.* κ = Cohen's kappa coefficient. Bold values indicate almost perfect agreement (κ ≥ 0.80); Sensitivity = true positive rate; Specificity = true negative rate; Mean Time = mean processing time in seconds; SD = standard deviation. Sample size: n = 500 cases (250 positive, 250 negative).

**Table 4.** Model Performance on Opioid Identification Benchmark

| Model Configuration | κ | Sensitivity | Specificity | Mean Time (SD) |
|---|---|---|---|---|
| Qwen3-0.6B Standard | 0.51 | 0.88 | 0.87 | 0.14 (0.14) |
| Qwen3-0.6B Reasoning | 0.64 | 0.94 | 0.88 | 1.32 (0.49) |
| Qwen3-1.7B Standard | 0.31 | 0.66 | 1.00 | 0.18 (0.05) |
| Qwen3-1.7B Reasoning | **0.87** | 0.95 | 0.99 | 2.40 (1.71) |
| Qwen3-4B Standard | **0.80** | 0.90 | 1.00 | 0.32 (0.10) |
| Qwen3-4B Reasoning | **0.96** | 0.99 | 0.99 | 3.18 (1.68) |
| Qwen3-8B Standard | **0.90** | 0.97 | 0.97 | 0.46 (0.13) |
| Qwen3-8B Reasoning | **0.96** | 0.98 | 1.00 | 5.15 (3.09) |
| Qwen3-14B Standard | **0.90** | 0.95 | 1.00 | 0.73 (0.21) |
| Qwen3-14B Reasoning | **0.95** | 0.98 | 1.00 | 6.44 (3.37) |
| Qwen3-32B Standard | **0.96** | 0.98 | 1.00 | 1.62 (0.50) |
| Qwen3-32B Reasoning | **0.96** | 0.98 | 1.00 | 13.10 (3.24) |
| Qwen3-30B-A3B Standard | 0.72 | 0.86 | 1.00 | 0.62 (0.23) |
| Qwen3-30B-A3B Reasoning | **0.96** | 0.98 | 1.00 | 4.02 (1.65) |
| gpt-oss-20b Low | **0.96** | 0.98 | 1.00 | 0.90 (0.40) |
| gpt-oss-20b Medium | **0.97** | 0.99 | 1.00 | 1.10 (0.96) |
| gpt-oss-20b High | **0.98** | 0.99 | 1.00 | 1.39 (1.61) |

*Note.* κ = Cohen's kappa coefficient. Bold values indicate almost perfect agreement (κ ≥ 0.80); Sensitivity = true positive rate; Specificity = true negative rate; Mean Time = mean processing time in seconds; SD = standard deviation. Sample size: n = 500 cases (250 positive, 250 negative).